\documentstyle[aps,epsf]{revtex}

\begin{document}

\title{Statistical theory of hierarchical avalanche ensemble}

\author{Alexander I. Olemskoi}
\address{Physical Electronics Department, Sumy State University\\
2, Rimskii-Korsakov St., 244007 Sumy UKRAINE\\
E-mails: Olemskoi@ssu.sumy.ua; Alexander@olem.sumy.ua}

\maketitle

\begin{abstract}
The statistical ensemble of avalanche intensities is
considered to investigate diffusion in ultrametric space of
hierarchically subordinated avalanches. The stationary
intensity distribution and the steady-state current are obtained.
The critical avalanche intensity needed to
initiate the global avalanche formation is calculated depending
on noise intensity. The large time asymptotic for the probability of
the global avalanche appearance is derived.

\end{abstract}


\section{Introduction}\label{sec1}

In recent years considerable study has been given to the theory
of self-organized criticality (SOC) that explains avalanche dynamics
in vast variety of systems type of sandpile model \cite{1},
intermittency in biological evolution \cite{2},
earthquakes and propagation of forest-fires, depinning transitions
in random medium and so on (see \cite{3}).
Without any special tuning of their external parameters, such systems evolve
to a critical state that displays universality
in both spatiotemporal behaviour and
avalanche distribution over size $s$ and intensity $f$ .
So, the latter is given by homogeneous function
\begin{equation}
P(s,f)=s^{-\tau}g(x),\qquad x\equiv s/s_c,
\label{1}
\end{equation}
where a cutoff size $s_c$ is determined as
\begin{equation}
s_c\sim (f_c-f)^{-1/\sigma},
\label{2}
\end{equation}
$f_c$ is a critical intensity, $\tau$, $\sigma$ are critical exponents.
Such relations are obtained as result
of making use of scaling-type arguments supplemented
with extensive computer simulations \cite{4}
as well as within framework of the theory of branching processes \cite{5}.
These methods allow to find the magnitudes of the parameters
$f_c$, $\tau$, $\sigma$,
but as to the function $g(x)$ in Eq.(\ref{1}),
it is known only that this function is a monotonically falling down.

This work is devoted to determination of the scaled function $g(x)$
within the framework of
statistical theory that deals with avalanche ensemble in the course
of SOC progressing.
The cornerstone of our approach is hierarchical representation
of the avalanche ensemble that is embedded into ultrametric space
corresponding to hierarchical landscape of the system energy \cite{6}.

The paper is organized as follows.
Sec.\ref{sec2} deals with the
theory of hierarchical coupling between elementary avalanches arising
from the complexity of the phase space landscape  of self-organized
system. The time dependence of the probability of the global
avalanche appearance is studied in Sec.\ref{sec3}.
Sec.\ref{sec4} contains discussion of obtained results.

\section{Intensity distribution in hierarchical avalanche
ensemble}\label{sec2}

Let the maximum number of
hierarchically subordinated avalanches $N$ be
on the bottom hierarchical level $s=0$ where the
avalanche intensity equals $f$. Correspondingly,
there is the only avalanche of the intensity $F\gg f$ on the top
level $s=s_{c}\gg 1$. The problem is to find the dependencies
$N(s)$ and $F(s)$ that define the distribution of avalanche number and
their intensity over hierarchical levels $s\in [0,s_{c}-1]$.

The first part of the problem can be approached in terms of geometry
by representing the avalanche ensemble as a hierarchical Cayley tree
\cite{7,8}. The basic types of the trees are shown in Fig.1: regular
tree with integer branching ratio $j$, regular Fibbonachi tree with
fractional one $j=\tau\approx 1.618$, degenerate tree with the only
branching node per level and the tree of our primary concern
-- irregular tree. Let $k$ be the numbering index for the levels, so
that $k$ increases from the top level to the bottom one. The
variable
\begin{equation}
s=s_{c}-k
\label{17}
\end{equation}
then defines the distance in the ultrametric space \cite{9}.
Geometrically, objects of this space correspond to the nodes of the
bottom level ($k=s_{c}$) of a Cayley tree. Since the distance between
the nodes is defined by the number of steps to a common ancestor, the
distance is eventually the level number (\ref{17}), counted from the bottom.

As it can be seen in Fig.1a, in the simplest case of regular tree
with integer branching ratio $j$ the number of avalanches
$N_{k}=j^{k}$ exponentially decays to zero with the distance
$s$ between them:

\begin{equation}
N(s)=N \exp{(-s\ln j)},\qquad N\equiv j^{s_{c}}.
\label{18}
\end{equation}
In Eq.(\ref{18}) the equality (\ref{17}) is used and the avalanche
number $N$ is related to the total number of levels $s_{c}$. For the
Fibbonachi tree (see Fig. 1b), where
$N_{k}=q\tau^{k},\,q\approx 1.171,\,\tau\approx 1.618$ \cite{8}, we
have
\begin{equation}
N(s)=N \exp{(-s\ln \tau)},\qquad N\equiv q\tau^{s_{c}}.
\label{19}
\end{equation}
When Eq. (\ref{19}) is compared with Eq. (\ref{18}), it is clear that the
exponential decay remains unaltered in the case of fractional
branching ratio and characterizes the regularity of tree.

For the degenerate tree (see Fig.1c) $N_{k}=(j-1)k+1$ and
Eq.(\ref{17}) provides the following linear dependence
\begin{equation}
N(s)=N-(j-1)s,\qquad N\equiv (j-1)s_{c}+1.
\label{20}
\end{equation}
It can be shown that in the case of irregular tree, displayed in
Fig.1d, the power law dependence is realized:
\begin{equation}
N_{k}=k^{a},\qquad a>1.
\label{21}
\end{equation}
Indeed, the latter can be regarded as an intermediate case between the
exponential Eqs.(\ref{18}), (\ref{19}) and linear Eq.(\ref{20})
obtained for the limiting cases of regular and degenerate
trees, respectively.
Formally, the approximation (\ref{21}) means that a function $N(x)$
defined on the self-similar set of hierarchically subordinated
avalanches is homogeneous, $N(kx)=k^{a}N(x)$.
It is convenient to rewrite Eq.(\ref{21}) in
term of the distance:

\begin{equation} N_{k}=N(1-s/s_{c})^{a},\qquad N\equiv
s_{c}^{a},\qquad a>1.
\label{22}
\end{equation}

Now, let us define $F_{k}$ as total intensity of avalanches on the $k$-th level.
If one considers the value $F_k$ as an effective density of some particles
and the level number $k$ as a coordinate,
then corresponding density of hierarchical current can be taken in the
Onsager-type form:
\begin{equation}
j_{k}=-D(F_{k})~\frac{\text{d}F_{k}}{\text{d}k}.
\label{23}
\end{equation}
Here, within the multiplicative noise approach,
the effective diffusion coefficient has form \cite{9a}
\begin{equation}
D(F)=DF^{-\alpha}
\label{24}
\end{equation}
to depend on the constant $D>0$ and the exponent $\alpha$.

The basic
assumption of this section is that the total current $J$ of all avalanches
at given hierarchical level is independent on the level number $k$:
\begin{equation}
j_{k}N_{k}=const\equiv J.
\label{25}
\end{equation}
Inserting Eqs.(\ref{22})-(\ref{24}) into Eq.(\ref{25}) gives the
total avalanches intensity on the level $k$
\begin{equation}
F_{k}=Fk^{-b},\quad
F^{1-\alpha}\equiv {1-\alpha\over a-1} {J\over D},\quad
b\equiv{a-1\over 1-\alpha}>0
\label{26}
\end{equation}
normalized by the maximum value $F\equiv F_{k=1}$.
Introducing the distance (\ref{17}), we obtain
\begin{equation}
F(s)=f(1-s/s_{c})^{-b},
\label{27}
\end{equation}
where the intensity at the bottom level $s=0$ is
\begin{equation}
f\equiv Fs_{c}^{-b}=FN^{-b/a}.
\label{28}
\end{equation}

After generalizing Eqs.(\ref{26}), (\ref{28}), the following scaling
relation can be assumed
\begin{equation}
F_{k}=N^{b/a}k^{-b}f_{k}
\label{29}
\end{equation}
where $f_{k}$ is a slowly varying function. According to
Eqs.(\ref{23})-(\ref{25}), this function obeys the
Landau-Khalatnikov equation:
\begin{equation}
\frac{\text{d}x}{\text{d}\kappa}=
-\frac{\partial V}{\partial x},
\label{30}
\end{equation}
where one denotes
\begin{equation}
\kappa\equiv\ln{k^{b}},\qquad x\equiv f_{k}/f_{c},\qquad
f_{c}^{1-\alpha}\equiv\left(J/bD\right)N^{-(a-1)/a}
\label{31}
\end{equation}
and the effective potential is introduced
\begin{equation}
V=\frac{x^{1+\alpha}}{1+\alpha}-\frac{x^{2}}{2}.
\label{32}
\end{equation}

As indicated in Fig.2, the potential $V$ reaches its maximum value
$V_{c}=(1-\alpha)/2(1+\alpha)$ at $x=1$ and decreases indefinitely at
$x>1$. So, in order to initiate the global avalanche formation, a low
intensity avalanche with $f<f_{c}$ at the bottom level needs to
penetrate the barrier $V_{c}$ of the potential (\ref{32}). It implies
fluctuation mechanism for the SOC regime progressing provided that
$x$ is a stochastic variable and we proceed with
Langevin-type equation derived from Eq.(\ref{30}) by adding a
Gaussian white noise to the right-hand side:

\begin{equation}
\frac{\text{d}x}{\text{d}\kappa}=
-\frac{\partial V}{\partial x}+\zeta,
\label{33}
\end{equation}

\begin{equation}
\langle\zeta\rangle=0,\qquad
\langle\zeta(\kappa)\zeta(\kappa^{\prime})\rangle=
2D\delta(\kappa-\kappa^{\prime}),
\label{34}
\end{equation}
where the noise intensity $D$ equals the diffusion coefficient
in Eq.(\ref{24}).

The usual way to study a set of solutions to the stochastic equation
(\ref{33}) is to introduce
distribution function $g(\kappa,x)$ associated with the probability
of solution's realization. It is known that $g(\kappa,x)$ obeys the
Fokker-Planck equation \cite{10}:
\begin{equation}
\frac{\partial g}{\partial \kappa}+
\frac{\partial j}{\partial x}=0,\qquad
j\equiv -g~\frac{\partial V}{\partial x}-
D~\frac{\partial g}{\partial x}.
\label{35}
\end{equation}
Since there is no current at the equilibrium state ($j=0$), the
corresponding distribution function of avalanche intensities
at the bottom level
\begin{equation}
g_{0}(x)\propto \exp{(-V(x)/D)}
\label{36}
\end{equation}
is dictated by the potential (\ref{32}). In the case of
non-equilibrium steady state the probability density $g$ does not
depend on the hierarchical level variable $\kappa$ and the current $j$
being constant, in compliance with the conservation law (\ref{25}),
can take a non-zeroth value.
From Eq.(\ref{35}) the stationary distribution then is
expressed in terms of the equilibrium distribution $g_{0}(x)$ and the
current $j$ \cite{11}:

\begin{equation}
\frac{g(f)}{g_{0}(f)}=\frac{j}{D}
\int\limits_{f/f_{c}}^{\infty}\frac{\text{d}x}{g_{0}(x)},
\label{37}
\end{equation}
where the boundary condition $g\to 0$ as $f\to\infty$ is taken into
account.

Given the intensity $f$ equation (\ref{37}) allows the current $j$ to be
found. In trying to do it, special consideration should be given to
the fact that the intensity $f$ is bounded from below, $f>G$,
by a gap $G$ that is inherent in
hierarchical ensemble of avalanches \cite{4}. Indeed, after merging of
avalanches within a hierarchical cluster of the size $s_{g}$, all
$s$, such that $s<s_{g}$, are appeared to be dropped out the
consideration as well as low intensities with $f<f(s_{g})\equiv G$
(see Fig.1). The expression for the current $j$ then can be derived from
Eq.(\ref{37}) with the second boundary condition $g(G)=g_{0}(G)$. The
result reads

\begin{equation} j=2DW\left\{ 1+\text{erf}\left[
\sqrt{\frac{1-\alpha}{2D}}
\left(1-\frac{G}{f_{c}}\right)
\right]\right\}^{-1},
\label{38}
\end{equation}
where the factor
\begin{equation}
W\propto\exp(-V_{c}/D),\qquad
V_{c}\equiv\frac{1}{2}~\frac{1-\alpha}{1+\alpha}
\label{39}
\end{equation}
gives the probability that fluctuation will surmount the barrier $V_{c}$ of
the potential (\ref{32}). Equation (\ref{38}) shows that in the case of
small gap, $G\ll f_{c}$, the current $j$ is about
$WD$, but the current is doubled under $G=f_{c}$. It can be understood
if we picture the effect of the gap as a mirror that reflects
diffusing particles at the point $f=G$: if $G\ll f_{c}$ a particle
penetrating the barrier can move along both directions, but in the
case of $G=f_{c}$ the mirror is placed at the point corresponding
to the top of the barrier and all particles go down the side where
the intensity $f$ grows indefinitely.

Given the current $j$ the stationary distribution function $g(f)$ is
defined by Eq.(\ref{37}), according to which,
$g(f)\approx g_{0}(f)$ in the subcritical region $f<f_{c}$, while in
the supercritical range $f\gg f_{c}$ we have $g_{0}(f)\gg g(f)$ due
to indefinite increase of $g_{0}(f)$. As far as the stationary
distribution is concerned, it can be derived from the current definition
(\ref{35}), where the last diffusion term is negligible for
supercritical intensities: $j\approx -(\partial V/\partial x)g$.
The result is that
the probability $g(f)$ remains almost unaltered $(g(f)\approx g(f_{c}))$
in the range from the critical value $f_{c}$ up to the boundary one $f_{g}$
and $g(f)\approx 0$ at $f\gg f_{g}$ (see \cite{11}).
The growth of $f_{g}$ is governed by the equation
\begin{equation}
\frac{\text{d}f_{g}}{\text{d}\kappa}=
D~\frac{f_{g}-f_{c}}{f_{g}^{2}}
\label{40}
\end{equation}
that is a counterpart of the known gap-equation presenting
the system behaviour below the critical value $f_c$ \cite{4}.

Since the above picture  is essentially statistical,
it enables the critical avalanche intensity $f_{c}$ for the transition
point to be found. Indeed, when the definition of the macroscopic current
$J$ in Eq.(\ref{25}) is compared to that of the microscopic current $j$
in Eq.(\ref{35}), it is apparent that they differ from one another
only by the factor $N^{(a-1)/a}\equiv s_{c}^{a-1}$ dependent on the
total number of avalanches $N$. On this basis,
the last expression of Eq.(\ref{31}) and Eq.(\ref{38}) at $G=0,\,D\ll
1$ give the desired result:

\begin{equation}
f_{c}=F\exp(-D_0/D),\qquad
D_0^{-1}\equiv 2(1+\alpha),
\label{41}
\end{equation}
where the pre-exponent factor $F$ determines the probability of the
barrier penetrating and cannot be calculated within the framework of
the presented approach. Equation (\ref{41}) bears a resemblance
to the well-known result of the superconductivity BCS theory for the temperature
of the phase transition and predicts the slow growth of the critical
intensity $f_{c}$ of elementary avalanche with the hierarchical
diffusion coefficient $D$ that plays the role of the parameter of
effective interaction.

\section{Kinetics of the global avalanche formation}\label{sec3}

Since the ensemble of hierarchically subordinated avalanches
represents a self-similar set, the probability distribution $P(s,f)$ of
avalanches in the course of SOC
process is a homogeneous function (\ref{1}),
where $g(s/s_c(f))\equiv g(f)$
is the stationary distribution of elementary avalanches
considered in the previous section.
Physically, Eq.(\ref{1}) implies that the total intensity $F$,
being measured by the scale $N^{b/a}$, equals the intensity
of an elementary avalanche $f$ in accordance with Eq.(\ref{28}).

In this section we are aimed to describe kinetics of the global
avalanche formation produced by virtue of the hierarchical coupling
between elementary avalanches. As it has been clarified
in Sec.\ref{sec2}, this process can be conceived of as diffusion in
ultrametric space that makes the distribution (\ref{1}) mounted.
In order to find the conditional probability
$\overline{{\cal P}}(t)$
that no global avalanche will appear at time $t$ one has to
integrate over $s$ the distribution (\ref{1}) weighted with the
function
\begin{equation}
p_{s}(t)=\exp{(-t/t(s))},\qquad
t(s)=t_{0}\exp{(F(s)/D)}
\label{43}
\end{equation}
descriptive of Debay relaxation with the time $t(s)$ governed by
the barrier height $F(s)$ ($t_0$ is a microscopic time).
By using the steepest descent method, it
is not difficult to derive the late time ($t\to\infty$) asymptotic
formula
\begin{equation}
\overline{{\cal P}}(t)=\left(\frac{f}{D}\right)^{\tau/b}
\left[1-\left(\frac{D}{f}
\ln{\frac{t}{t_{ef}}}
\right)^{-1/b}
\right]^{-\tau},\qquad
t_{ef}\equiv\frac{\tau}{b}\left(\frac{f}{D}\right)^{1/b}t_{0}.
\label{44}
\end{equation}
This equation has been obtained by assuming that the condition
$1\ll s_{m}\le s_{c}$ is met, where $s_{m}$ denotes the location of the
maximum of integrand and obeys the equation
\begin{equation}
\frac{D\tau}{bf}\frac{(1-y)^{1+b}}{y}=
\frac{t}{t_{0}}\exp{\left(-\frac{f}{D}(1-y)^{-b}\right)},\qquad
y\equiv \frac{s_{m}}{s_{c}}.
\label{45}
\end{equation}
Taking into consideration the scaling relation (\ref{2}) for the number of
hierarchical levels $s_{c}$, which is the cut-off parameter,
we readily come to the conclusion that the condition is satisfied provided

\begin{equation}
f-f_{c}\ll f,\qquad t\gg t_{ef}
\exp{\left(\left(f_{c}/D\right)^{-1/b}-1 \right)^{-b}}.
\label{47}
\end{equation}
So, the intensity $f$ in
Eqs.(\ref{44}),(\ref{45}) can be replaced by the critical value $f_{c}$.
In accordance with Eq.(\ref{44}), the probability
${\cal P}(t)\equiv 1-\overline{\cal P}(t)$
of the global avalanche appearance logarithmically increases
in time up to the value ${\cal P}=1-(f_{c}/D)^{\tau/b}$.
In order for the probability ${\cal P}$ to be non-negative, the factor
$F$ in Eq.(\ref{41}) must be equal $F_{0}\equiv (e/2)(1+\alpha)^{-1}$,
whereas the effective diffusion coefficient $D$ must be bounded from above by the value
$D_{0}\equiv (1/2)(1+\alpha)^{-1}$.

\section{Discussion}\label{sec4}

According to the presented picture,
the initiated elementary avalanches form statistical ensemble of
hierarchically subordinated objects, characterized by intensity
$f$ and distance in ultrametric space $s$
(the latter corresponds avalanche size \cite{4}).
Since the global avalanche
formation is caused by effective diffusion in the space, then, similar
to Brownian particle with coordinate $f$ at time $s$, the ensemble
can be described by Langevin equation (\ref{33}) subjected to the
noise Eq.(\ref{34}) with $D$ being the effective diffusion
coefficient and corresponding to the Fokker-Planck equation (\ref{35}).
The stationary intensity distribution and the steady-state current
are given by Eqs.(\ref{37}), (\ref{38}). The condition of current
conservation Eq.(\ref{25}) yields the avalanche intensity distribution
(\ref{27}) over hierarchical clusters in the ultrametric
space. The ensemble of elementary avalanches, being weakly
dependent on $s$, is governed by the effective potential (\ref{32})
that reaches its maximum at the critical intensity (\ref{41}) (see
Fig.2). So, the global avalanche generation requires
supercritical elementary avalanche intensity, $f>f_{c}$, to surmount
the barrier $V_{c}$ with the characteristic time (cf. Eq.(\ref{39}))
\begin{equation}
T\approx t_{0}\exp{(V_{c}/D)},\qquad
V_{c}\equiv\frac{1}{2}~\frac{1-\alpha}{1+\alpha}.
\label{49}
\end{equation}
This picture bears some resemblance with the formation process of
supercritical embryo in the theory of the first-order phase
transitions \cite{11}, where in the course of transformation the next
stage is the diffusion growth of the embryo.
Analogously, in the case under
consideration the above growth implies an increase of the
supercritical avalanche in intensity $F(s)$, Eq.(\ref{27}), due to
the diffusion growth of hierarchical cluster in ultrametric space.
As a result of the total cluster formation, we have
the logarithmically slow large time asymptotic for the probability
of the global avalanche appearance:
\begin{equation}
{\cal P}(t)=1-\overline{\cal P}
\left[1-\overline{\cal P}^{1/\tau}\left(
\ln{\frac{t-T}{t_{ef}}}
\right)^{-1/b}
\right]^{-\tau},\qquad
t_{ef}\equiv(\tau/b)
\overline{\cal P}^{1/\tau}t_{0},
\label{50}
\end{equation}
where time $t$ is
counted from the instant  $T$, Eq.(\ref{49}), and
$\overline{\cal P}$ is the maximum
probability that no global avalanche will occur
\begin{equation}
\overline{\cal P}=\left(
\frac{D_{0}}{D}
\right)^{\tau/b}\exp{\left[
-\frac{\tau}{b}\left(
\frac{D_{0}}{D}-1
\right)
\right]}.
\label{51}
\end{equation}
From Eq.(\ref{51}) the probability is determined by the ratio of
the noise intensity $D$ (see Eq.(\ref{34})) and its maximum value
$D_{0}=(1/2)(1+\alpha)^{-1}$. The key point is that the maximum
probability ${\cal P}\equiv 1-\overline{\cal P}$ of the global
avalanche appearance is completely suppressed under high intensity
variance in ensemble of elementary avalanches (see Fig.3).

\vspace{2 cm}

\begin{center}
{\bf FIGURE CAPTIONS}
\end{center}

\begin{figure}
\caption{
Different types of hierarchical trees
(the level number is indicated at left, corresponding number of
nodes -- at right): a) regular tree with $j=2$; b) Fibonacci tree;
c) degenerate tree with $j=3$; d) irregular tree.
}
\label{fig1}
\end{figure}

\begin{figure}
\caption{
The effective potential (18) as a function of
$f/f_{c}$
at $\alpha=0.1$.
}
\label{fig2}
\end{figure}

\begin{figure}
\caption{
The dependence of
the maximum
probability ${\cal P}\equiv 1-\overline{\cal P}$ of the global
avalanche appearance on the intensity
variance $D$ in ensemble of elementary avalanches.
}
\label{fig3}
\end{figure}

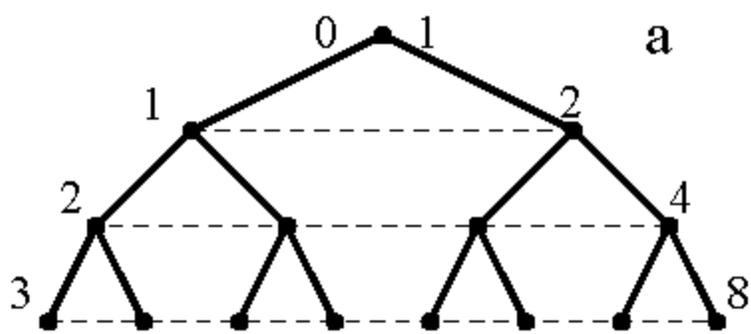
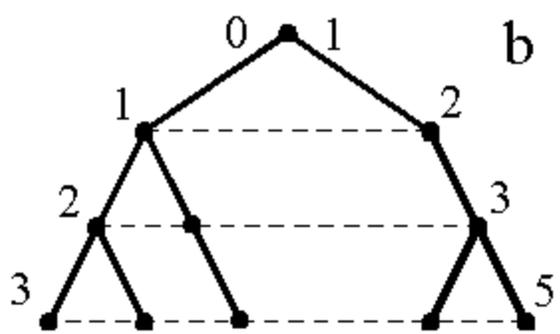
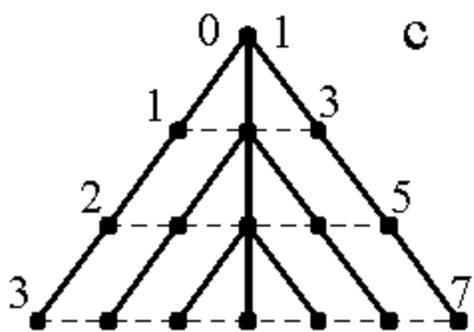
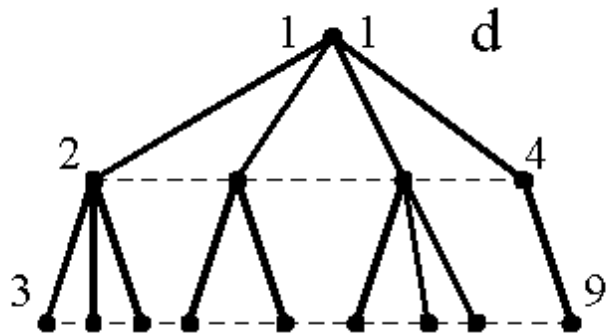
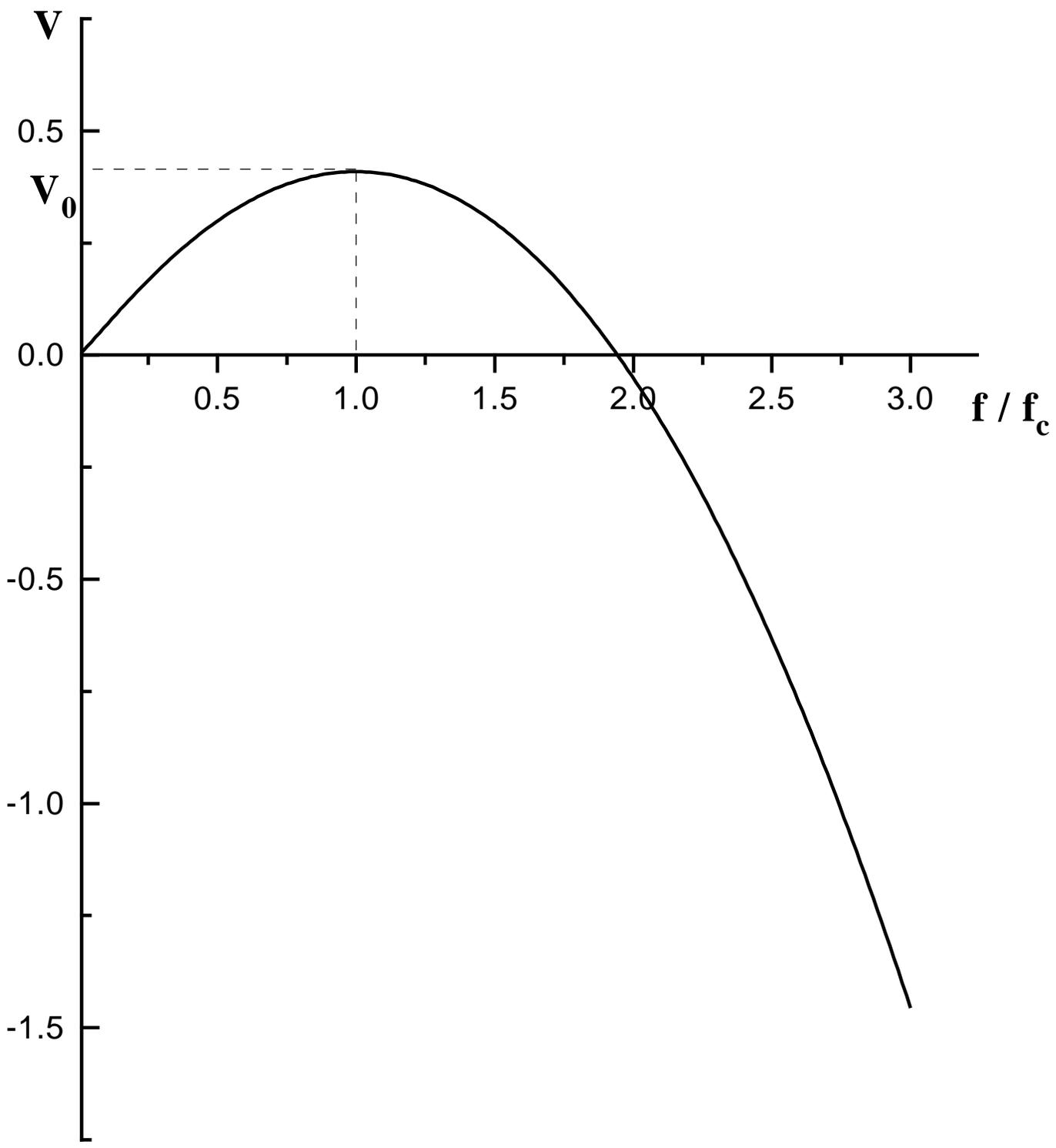
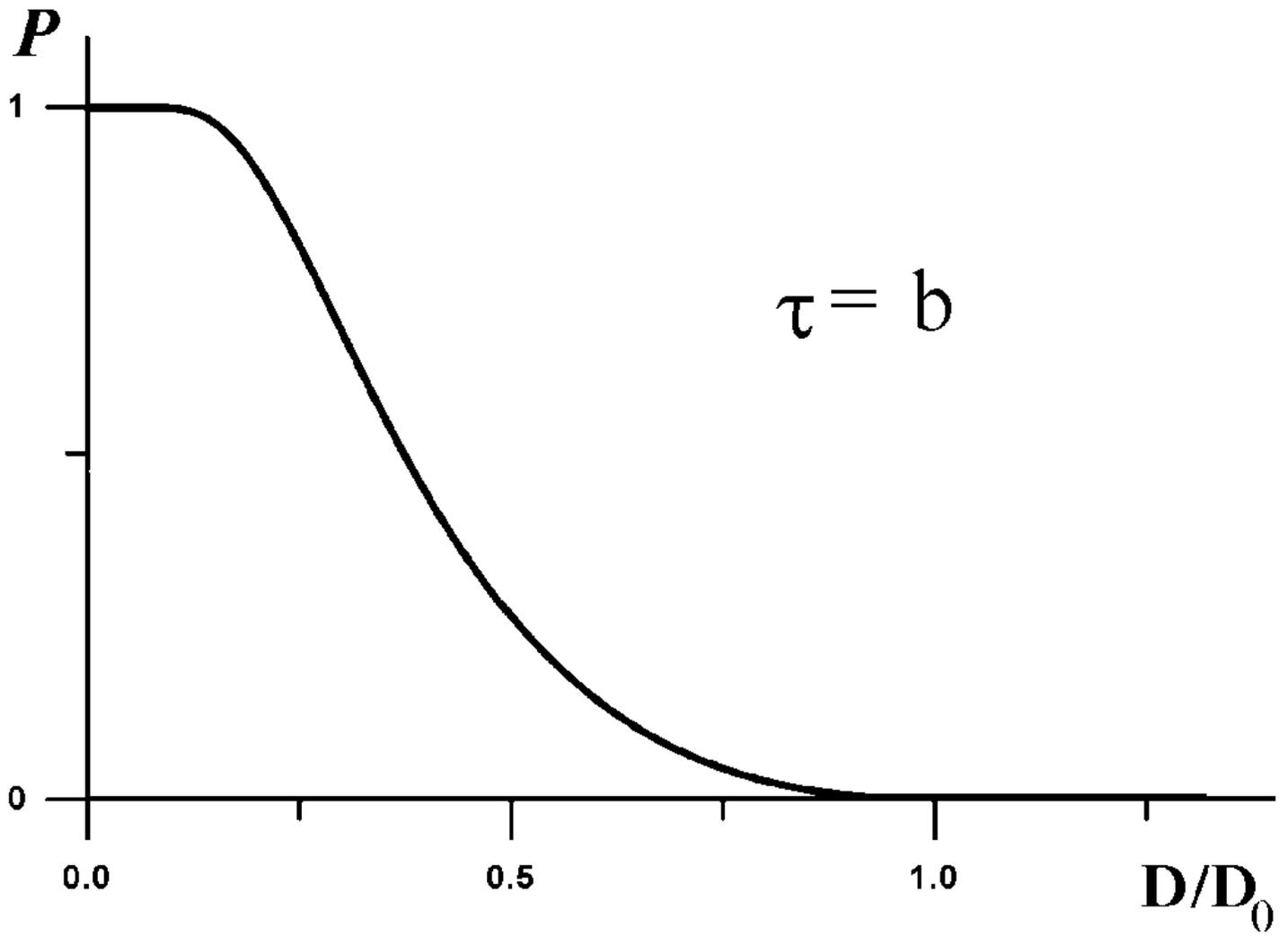
\end{document}